\documentclass[iop,apjl]{emulateapj}
\usepackage{graphicx}
\usepackage{dcolumn} 
\usepackage{bm}      
\usepackage{color}   
\usepackage{amsmath}
\usepackage{natbib}

\newcommand{\bea}{\begin{eqnarray}}
\newcommand{\eea}{\end{eqnarray}}

\newcommand{\Sobs}{S_{\rm obs}}
\newcommand{\nobs}{n_{\rm obs}}

\newcommand{\mumin}{\mu_{\rm min}}
\newcommand{\Msun}{M_{\odot}}
\newcommand{\Lsun}{L_{\odot}}

\newcommand{\simgt}{\lower.5ex\hbox{$\; \buildrel > \over \sim \;$}}
\newcommand{\simlt}{\lower.5ex\hbox{$\; \buildrel < \over \sim \;$}}

\shorttitle{Submm. Galaxy Counts and Magnification by Clusters}
\shortauthors{M. Lima, B. Jain, M. Devlin \& J. Aguirre}

\begin{document}

\title{Submillimeter Galaxy Number Counts and Magnification by Galaxy Clusters}
\author{Marcos Lima, Bhuvnesh Jain, Mark Devlin, and James Aguirre}
\email{mlima@sas.upenn.edu}
\affiliation{
Department of Physics \& Astronomy, University of Pennsylvania, Philadelphia PA 19104
}

\begin{abstract}
\baselineskip 11pt
	We present an analytical model which reproduces measured
	galaxy number counts from surveys in the wavelength range of
	500 $\mu$m to 2 mm. The model involves a
	single high-redshift galaxy population with a Schechter luminosity function which has
	been gravitationally lensed by galaxy
	clusters in the mass range $10^{13}$ to $10^{15}\Msun$.   
        This simple model reproduces both the low flux and
        the high flux end of the number counts reported
	by the BLAST, SCUBA, AzTEC and the SPT surveys. In particular, 
	our model accounts for the most luminous galaxies detected by SPT 
	as the result of high magnifications by galaxy clusters 
        (magnification factors of 10-30).
	This interpretation implies that submillimeter and millimeter 
        surveys of this population may prove to be a useful 
	addition to ongoing cluster detection surveys. The model also implies
	that the bulk of submillimeter galaxies detected at wavelengths
	larger than 500 $\mu$m lie at redshifts greater than 2. 
\end{abstract}



\section{Introduction}
\label{sec:intro}

Over the last decade, submillimeter (submm) surveys have yielded significant 
advances in our understanding of the galaxy population responsible for 
the high-redshift component of the cosmic infrared background (CIB)
\citep{Fixetal96,SmaIviBla97,Hugetal98,Baretal98,Dweetal98,Fixetal98,Greetal04,Popetal06,Copetal06,Devetal09}.
With typical far-infrared (FIR) luminosities $> 10^{12} \Lsun$, submm galaxies 
are presumed to be the high-redshift counterparts to (ultra) luminous infrared galaxies (LIRGs,
ULIRGs). The high luminosity of these galaxies is the result of 
star formation rates of 100--1000 $\Msun$\,yr$^{-1}$.
Approximately half of these galaxies are located at $1.9 \lesssim z \lesssim 2.9$
\citep{Chaetal05,Areetal07}, dominating the total star formation rate at
this epoch \citep{Peretal05, Micetal09}. 

One way to express the results of submm surveys is through
number counts of galaxies as a function of flux for each observed
wavelength.  The shape of these counts has been interpreted as arising from
 different populations of galaxies whose characteristics 
 evolve over cosmic time \citep{Lagetal03,Lagetal04,Peaetal09,LeBetal09}.
These empirical models have successfully reproduced the counts.
However, they may be masking a simpler explanation for the
departure of the counts from a Schechter distribution at the high-flux
end: magnification due to high redshift galaxy clusters and groups 
\citep[e.g.][]{Bla96,Peretal02,Negetal07}. 

Millimeter wavelength surveys have also aimed at detecting galaxy clusters 
via the Sunyaev Zel'dovich (SZ) effect \citep{Hinetal08, Caretal09}; the first 
results, including CMB power spectra and cluster 
catalogs, have been released recently 
\citep{Fowetal10, Staetal09, Vanetal10}.  The number of detected clusters
remains relatively low, primarily due to the low value of $\sigma _8$.  However,
other effects could be reducing the sensitivity of the surveys
\citep[e.g.][]{LimJaiDev09, Limetalinprep}.

The South Pole Telescope (SPT) has measured number counts of dusty 
galaxies at wavelengths $\lambda=1.4$ mm and $2.0$ mm over an area of 
$87$ deg$^2$ \citep{Vieetal09}.
The observed numbers at the bright end 
are higher than expected: these galaxies are either 
at high redshifts and intrinsically exceptionally luminous, or have been magnified by 
gravitational lensing, or are simply at much lower redshifts than the bulk of the 
population of submm galaxies. The latter possibility is disfavored by the lack of detected
counterparts in other surveys that probed the low redshift population 
\citep{Vieetal09}. The possibility that these galaxies are at high redshifts and 
intrinsically bright would require them to be far more luminous than an underlying
Schechter-like luminosity function would permit. Thus the favored explanation is that
they have typical luminosities for high-$z$ galaxies, but have been magnified by foreground 
galaxies or clusters. In fact, lensing of high-redshift background 
submm galaxies has been observed in a number of systems 
\citep{SmaIviBla97,Smaetal02,Wiletal08,Rexetal09,Gonetal09,Swietal10}. 

In this {\it Letter}, we explore the possibility that the existing observed galaxy number
counts over a wide range of wavelengths can be reproduced by a {\it single} population of 
galaxies at high-redshift.  Foreground galaxy groups and clusters
gravitationally lenses the background submm population
\citep{LimJaiDev09} and leads to significant enhancements of the high-flux end to
the galaxy counts. In \S~\ref{sec:lensing} we describe the lensing magnification formalism, 
which we then apply to a high-$z$ galaxy population and present results in 
\S~\ref{sec:results}. We discuss implications 
for high-$z$ galaxies and the cluster searches in \S~\ref{sec:discussion}.

\begin{figure}
\resizebox{88mm}{!}{\includegraphics[angle=0]{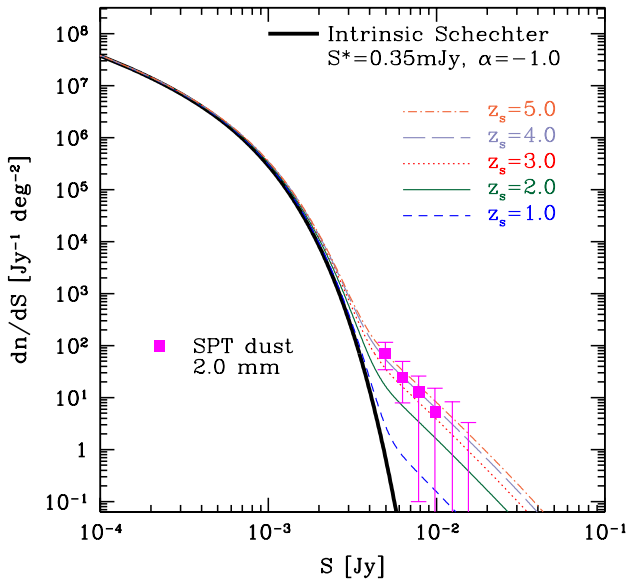}}
   \caption{Intrinsic and lensed number counts $dn/dS$ for a Schechter function describing 
galaxies at different redshifts. Also shown are the observed counts for SPT 
{\it dusty} submm galaxies at $\lambda=2.0$~mm, after removal of low-redshift galaxies with 
IRAS counterparts. 
}
 \label{fig:dndS_SPT_lens}
\end{figure}
%

Throughout, we use a fiducial cosmology for a flat universe with parameter 
values based on the results of the Wilkinson Microwave Anisotropy Probe 
fifth year data release \citep[WMAP5,][]{Kometal09}. 
The cosmological parameters (and their values) 
are the normalization of the initial curvature spectrum 
$\delta_\zeta (=2.41\times 10^{-4})$ at $k=0.02$ Mpc$^{-1}$
(corresponding to $\sigma_8=0.8$), its tilt
$n (=0.96)$, the baryon density relative to critical
$\Omega_bh^2 (=0.023)$, the matter density
$\Omega_{\rm m} h^2 (=0.13)$, and two dark energy parameters: its density
$\Omega_{\rm DE} (=0.74)$ and equation of state
$w(=-1)$, which we assume to be constant. 
Since lensing effects depend on the abundance of dark matter 
halos, which is exponentially sensitive to linear matter perturbations, 
we also consider changes in $\sigma_8$ consistent with the WMAP5 errors
of $\Delta \sigma_8 \approx 0.03$. Our central value and uncertainty for $\sigma_8$ 
is  consistent with the WMAP7 analysis \citep{Kometal10}.

\section{Number Counts with Lensing Magnification}
\label{sec:lensing}

In recent papers  \citep{LimJaiDev09, JaiLim10} we have presented a halo model 
for calculating the effects of lensing magnification by galaxy groups and clusters. 
Here we specialize to the case of
steep galaxy counts at high redshifts, where lensing effects are quite dramatic.  
We assume a Schechter function \citep{Sch76} for the intrinsic number density 
distribution of a population of galaxies
\bea
\frac{dn}{dS} = \frac{n^*}{S^{*}} \left(\frac{S}{S^*}\right)^{\alpha} e^{-S/S^{*}}\,,
\label{eq:Schechter}
\eea
where $n^*$, $S^*$ and $\alpha$ are free parameters. 
Due to lensing magnification by intervening halos, the intrinsic 
 $dn/dS$ is changed to its observed counterpart as
\begin{eqnarray}
\frac{d\nobs(\Sobs)}{d\Sobs} &=& \int d\mu \frac{P(\mu)}{\mu} \frac{dn}{dS}\left(\frac{\Sobs}{\mu}\right) \,,
\label{eq:dnobsdSobs}
\end{eqnarray}
where $\mu$ is the lensing magnification and $P(\mu)$ is its probability for a given galaxy 
population at redshift $z_s$.  
Conditional probabilities quantify the effects of different
magnification ranges on the observed flux density $\Sobs$. 
The integrand of Eq.~\ref{eq:dnobsdSobs} defines the probability $P(\mu|\Sobs)$
\begin{eqnarray}
P(\mu|\Sobs) &=& \left(\frac{d\nobs(\Sobs)}{d\Sobs}\right)^{-1}
                 \frac{P(\mu)}{\mu} \frac{dn}{dS}\left(\frac{\Sobs}{\mu}\right) \,,
\label{eq:PmuSobs}
\end{eqnarray}
which can be interpreted as the relative contribution of a given $\mu$
to the total $d\nobs/d\Sobs$ at $\Sobs$ \citep{PacScoCha08}. 
Similarly $P(\mumin|\Sobs)= \int_{\mumin}^{\infty} d\mu \ P(\mu|\Sobs) $
measures the integrated contribution from all $\mu> \mumin$.
The mean magnification at a given $\Sobs$ is defined as
\begin{eqnarray}
\langle \mu\rangle(\Sobs) &=& 
                           \int_0^{\infty} d\mu \ \mu \ P(\mu|\Sobs)\,.
\end{eqnarray}

The distribution $P(\mu)$ can be estimated either by ray-tracing on N-body simulations 
\citep[e.g.][]{Hiletal08},
or by semi-analytical methods \citep[e.g.][]{Peretal02,LimJaiDev09}, integrating halo 
contributions on the line of sight up to the source redshift
%
\begin{figure*}
\resizebox{88mm}{!}{\includegraphics[angle=0]{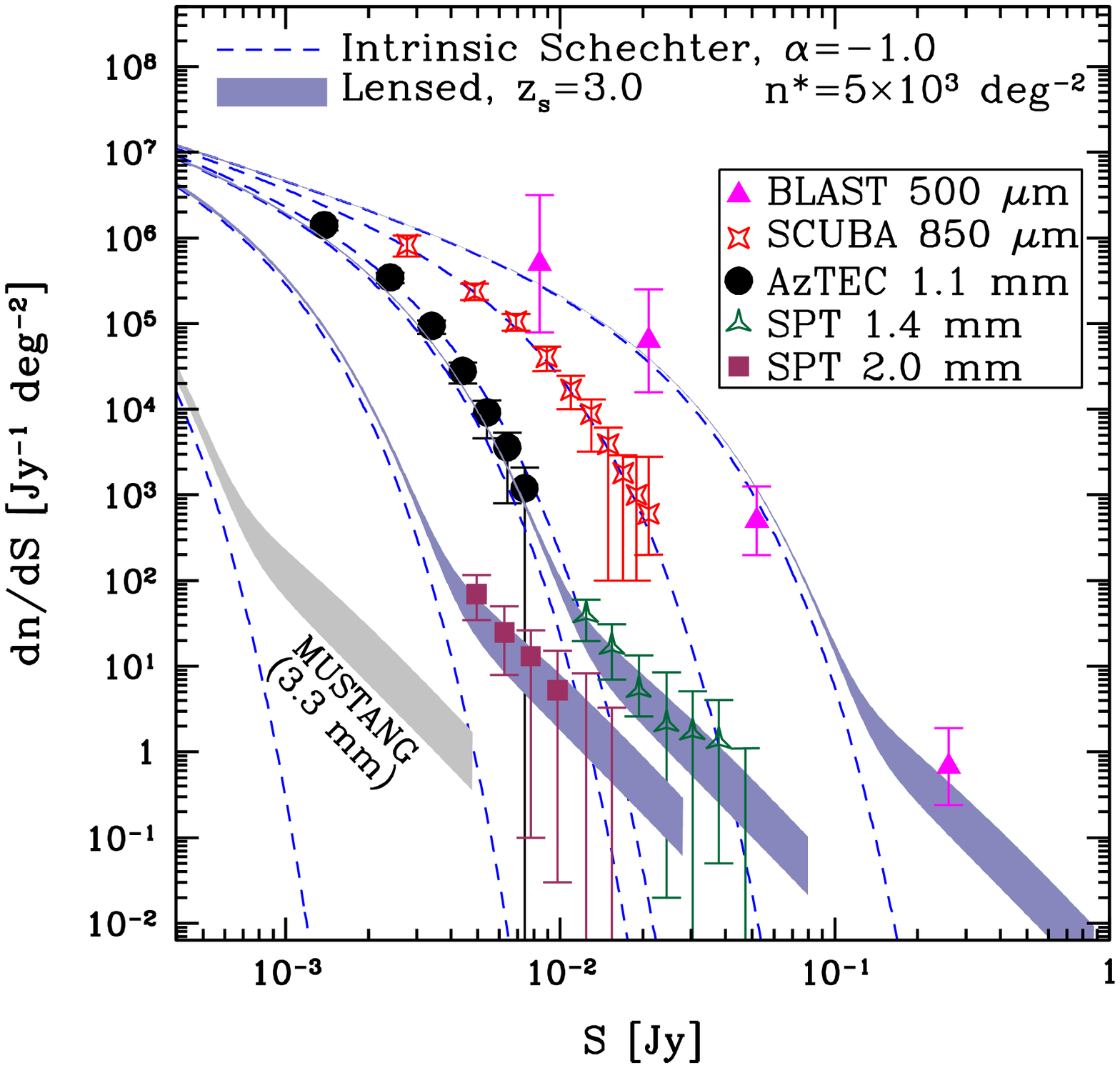}}
\resizebox{88mm}{!}{\includegraphics[angle=0]{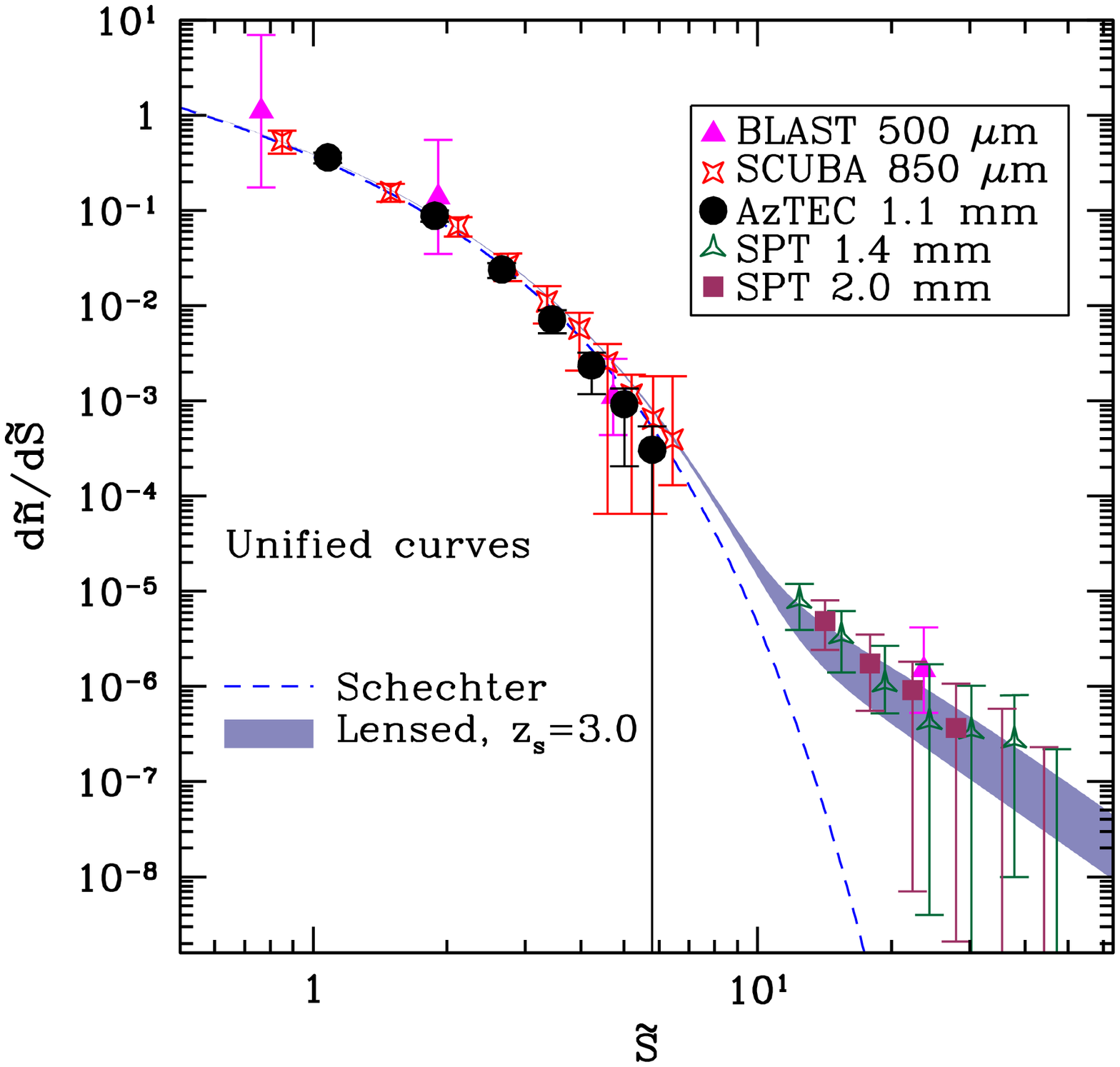}}
   \caption{({\it Left}) : Intrinsic and lensed $dn/dS$ for a Schechter function describing 
galaxies at $z_s=3.0$ and different wavelengths. Dashed lines indicate the intrinsic 
Schechter functions (with different $S^*$ values) and the dark shaded regions display the 
range of 
lensing predictions, as described in the text. 
Also shown are observed counts for BLAST at 
$\lambda=500$ $\mu$m, SCUBA at $\lambda=850$ $\mu$m, AzTEC at $\lambda=1.1$ mm and 
SPT at both $\lambda=1.4 $ mm and $2.0$ mm. 
The SPT counts are for {\it dusty} galaxies, after removal of galaxies with 
synchrotron emission and galaxies with IRAS counterparts. 
No similar removal has been applied to the BLAST data, which 
includes both high and low redshift galaxies. Notice that we do not display the lensing
predictions for SCUBA and AzTEC, since these data do not require lensing in the measured
fluxes.
A prediction for MUSTANG at $\lambda=3.3$ mm is also shown in the light shaded region.
({\it Right}) : Unified scaled curves showing $d\tilde{n}/d\tilde{S}$ and the various 
data points. The only parameter used in the scaling is $S^*$; its
values at the different  wavelengths are shown in
Fig.~\ref{fig:Sstar}. 
}
 \label{fig:dndS_all}
\end{figure*}
%
\begin{eqnarray}
P(>\mu)=\int_{0}^{z_s} dz_l \frac{D_A^2(z_l)}{H(z_l)} 
          \int_{M_{\rm th}}^{\infty} d\ln M
          \ \frac{dn(z_l,M)}{d\ln M}
          \ \Delta \Omega_{\mu} \,, \nonumber \\
\label{eq:Pmu}
\end{eqnarray} 
where $D_A$ is the angular diameter distance, $H$ is the Hubble parameter, 
$dn/d\ln M$ is the halo mass-function and 
$\Delta \Omega_{\mu}=\Delta \Omega_{\mu}(z_s,z_l,M)$ is the cross-section 
for magnifications larger than $\mu$ produced by halos of mass $M$ at 
redshift $z_l$ on sources at redshift $z_s$.
The integrand
\begin{eqnarray}
\frac{d^2P(>\mu)}{d \ln M \ dz_l}&=& \frac{D_A^2(z_l)}{H(z_l)} 
          \ \frac{dn(z_l,M)}{d\ln M}
          \ \Delta \Omega_{\mu} \,,
\label{eq:dPdM}
\end{eqnarray} 
gives the range of halo masses and redshifts contributing most to the 
probability of a specified minimum magnification $\mu$. 

In summary, Eqs.~\ref{eq:dnobsdSobs} and \ref{eq:Pmu} give the total effect on the counts, 
Eq.~\ref{eq:PmuSobs} indicates which 
magnifications contribute most to the given $\Sobs$ and Eq.~\ref{eq:dPdM} 
tells us which halo masses and redshifts contribute to a given magnification.

We use this halo-model $P(\mu)$ and correct it for a number of effects.
First, as described in \cite{LimJaiDev09}, we match our $P(\mu)$ at 
large magnifications to that of ray-tracing in dark matter simulations 
\citep{Hiletal08} by tuning the ellipticity of our halos. 
Next, we account for the effect of luminous matter, which can lead to higher densities via 
gas cooling, using the simulation results of \cite{Hiletal08b}. We also 
correct for the combination of finite source size and multiple image effects. 
Magnification effects, especially for large magnifications, are sensitive to the 
value of $\sigma_8$ since it affects the abundance of cluster halos. 
In the next section we discuss how we account for the uncertainties in our model 
by giving a range for our predictions. 

Our analytical calculation of $P(\mu)$ has some advantages over the 
approach of numerical simulations (we can easily study changes in source
redshift, $\sigma_8$ and the contribution from different halo masses
and redshift),  but it also has some limitations.  
We only use the one-halo term, which is accurate at the high magnifications
relevant for the effects considered here but overestimates the lensing 
contribution at $\mu\sim 1$. 
We do not include a distribution of ellipticities or halo
substructure, which can also increase magnification cross-sections. We
have instead tuned the average halo ellipticity to match the $P(\mu)$
measured in dark matter simulations of \cite{Hiletal08} 
(see \cite{LimJaiDev09} for a detailed discussion). 
And whereas we account for the effects of baryons observed 
in numerical simulations by \cite{Hiletal08b}, these authors note that
their simulation  still underestimate baryonic effects for 
halos of smaller masses as they do not predict sufficient numbers of
multiply imaged quasars. Finally, the effect of finite source size is
very uncertain given the lack of our knowledge  
about submm galaxies and its sensitivity to the precise caustic
structure of the lenses (e.g. \cite{Lietal05}).  
A proper inclusion of all these missing
effects would likely increase the magnification  
probabilities compared to our current model.

\section{Results}
\label{sec:results}


\begin{figure}
\resizebox{88mm}{!}{\includegraphics[angle=0]{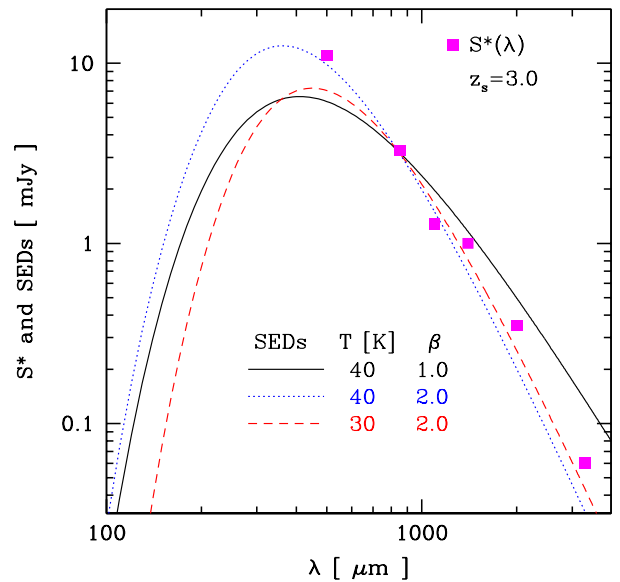}}
   \caption{ The flux $S^*$ in the Schechter function is shown for different
     surveys at various wavelengths (symbols). The three curves show 
the expected scaling for submm galaxy  
SEDs at $z_s=3.0$ for different values of spectral index $\beta$ and
     temperature $T$. The curves are normalized at the value of $S^*$ for 
$\lambda=850 \mu$m.
}
 \label{fig:Sstar}
\end{figure}

\begin{figure*}
\resizebox{88mm}{!}{\includegraphics[angle=0]{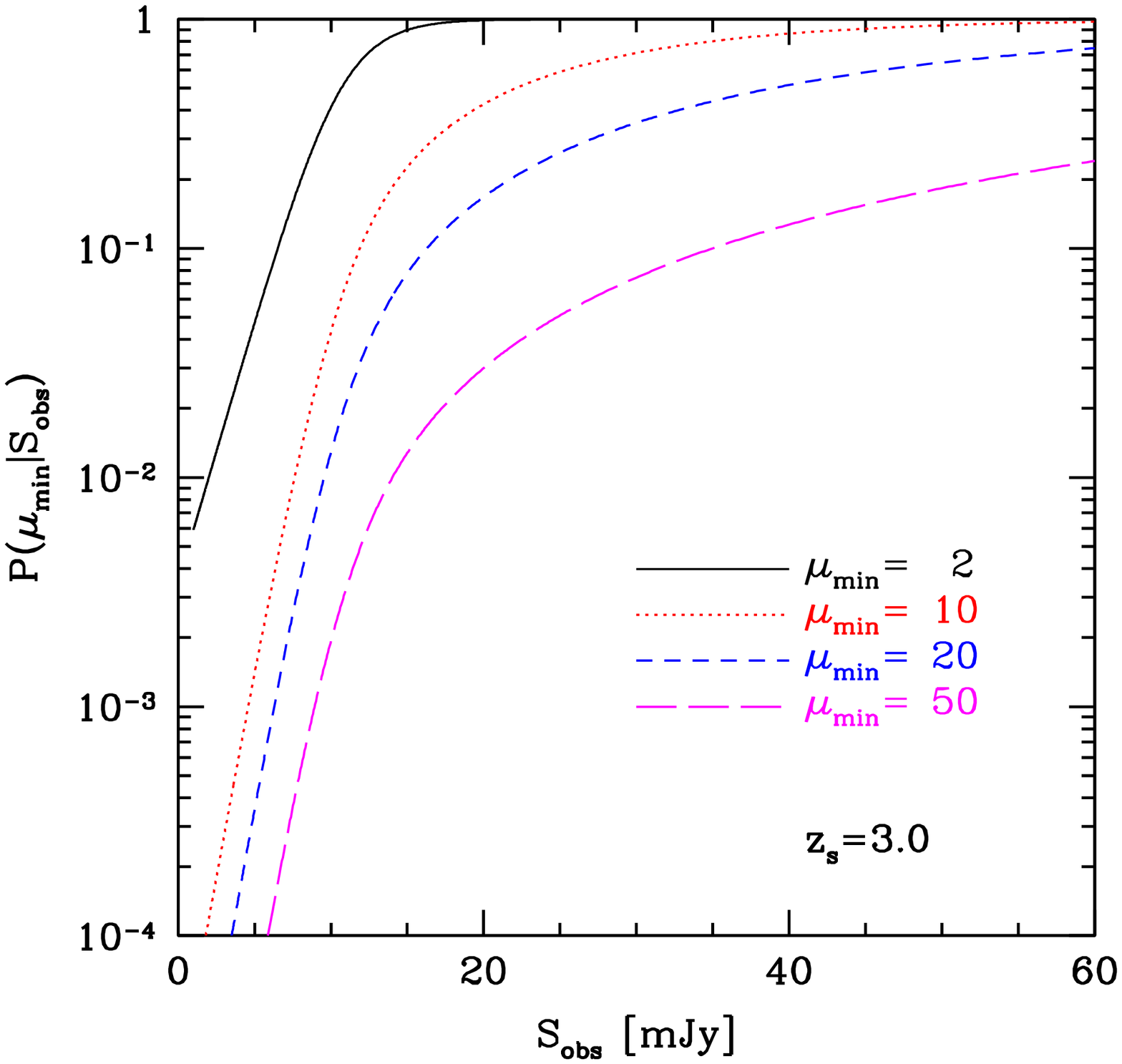}}
\resizebox{88mm}{!}{\includegraphics[angle=0]{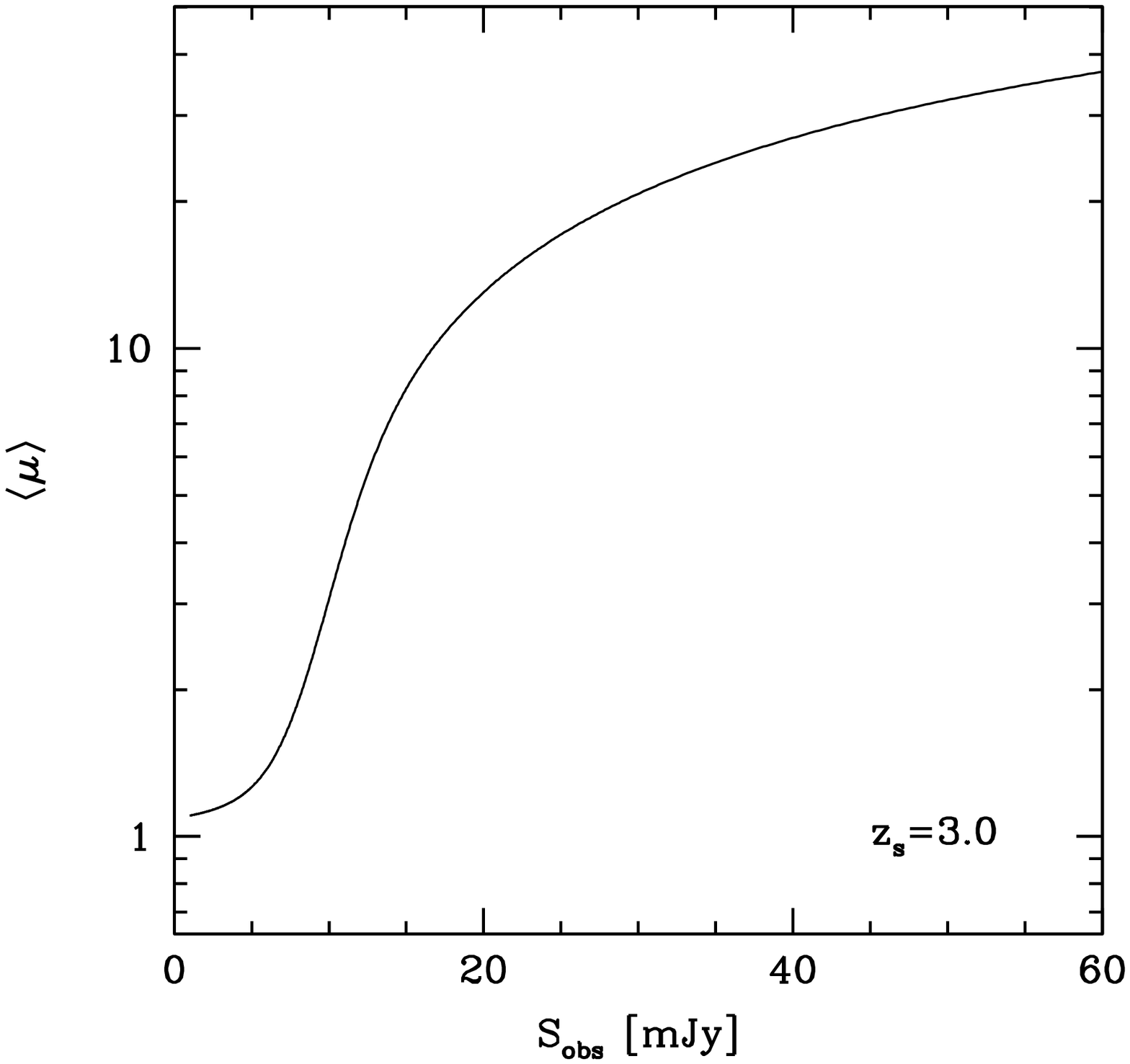}}
   \caption{ Probability $P(\mumin|\Sobs)$ of the minimum magnification $\mumin$, 
given an observed flux density $\Sobs$ (left panel), and the corresponding 
average magnification $\langle \mu \rangle$ (right panel). We assume a 
Schechter function describing galaxies at $z_s=3.0$ which, after lensing, 
predicts counts consistent with those of SPT dusty submm galaxies at $\lambda=1.4$~mm 
(see Fig.~\ref{fig:dndS_all}). 
}
 \label{fig:PmuminSobs.vs.S}
\end{figure*}

In Fig.~\ref{fig:dndS_SPT_lens} we illustrate the lensing effect on an intrinsic Schechter 
function distribution given by Eq.~\ref{eq:Schechter} for sources at different redshifts.
In all results presented here, we fix $\alpha=-1.0$ and $n^*=5 \times 10^3$~deg$^{-2}$. 
Changing to $\alpha=-1.5$ does not have a significant effect -- while it 
matches the faint end behavior of $dn/dS$ for the model of \cite{Lagetal04},  
we preferred to use $\alpha=-1.0$ since this fits better the 
number counts at shorter wavelengths.  
With these parameter values, our counts are lower than the model of \cite{Lagetal04} at 
all $S$, which also ensures that the total flux  
does not exceed the CIB \citep{Dweetal98, Fixetal98}.

Fig.~\ref{fig:dndS_all} illustrates our main results: 
we show predicted number counts that include lensing 
(gray bands), assuming galaxies at $z_s=3.0$, 
along with measured number counts from submm 
surveys at different wavelengths. As indicated in the panels, these are: BLAST 
at $\lambda=500$~$\mu$m \citep{Devetal09}, 
SCUBA at $\lambda=850$~$\mu$m \citep{Copetal06}, AzTEC at 1.1~mm \citep{Ausetal10} and 
SPT {\it dusty} submm galaxies at $\lambda=1.4$~mm and $2.0$~mm \citep{Vieetal09}. 
In all our results, the SPT number counts correspond to those of \cite{Vieetal09}, 
after {\it removing} both synchrotron emission galaxies as well as 
low-redshift galaxies that have matches with galaxies in the Infrared Astronomy Satellite survey 
\citep[IRAS,][]{Mosetal92,Fisetal95, Olietal96}. 
Predictions are shown by bands rather than curves to reflect the uncertainties in 
the model as discussed below. The intrinsic luminosity function is the Schechter function 
described above. All the data sets can be 
fit  by changing the single parameter $S^*$ once lensing magnification is included. 
This remarkable result implies that, 
within the measurement and theoretical uncertainties,  
a single high-$z$ population of galaxies is sufficient to describe all the observations. 
The high flux measurements of BLAST and SPT are fit by highly magnified galaxies -- 
if these counts were dominated by a population of a different galaxy type, it would be a
coincidence that their relative counts fit the same scaling with wavelength as the fainter 
(``normal'') population. 
Finally, we also show the prediction for the Multiplexed Squid TES
Array at 90 GHz (3.3mm) \citep[MUSTANG,][]{Masetal06} which has begun
operating on the Green Bank Telescope.

The lower bound for the predictions in  Fig.~\ref{fig:dndS_all}
uses $\sigma_8=0.77$ while the upper bound uses $\sigma_8=0.83$ -- these 
reflect the uncertainties in the WMAP results as discussed above. The enhancement
due to baryons is factored into these predictions following \cite{Hiletal08b}, though   
it is likely to be an underestimate of baryonic effects as discussed above in Section 2. 
The effect of source sizes and multiple imaging is uncertain; 
we simply assume that 
due to the finite source size the effective magnification is 
reduced by 50\% to 25\% (for the lower and upper bounds respectively)

The right panel of Fig.~\ref{fig:dndS_all} shows all points rescaled by plotting 
\bea
\frac{d\tilde{n}}{d\tilde{S}}=\tilde{S}^\alpha \ e^{-\tilde{S}}\,,
\eea
where 
$\tilde{n}=n/n^*$ and $\tilde{S}=S/S^*$.
In Fig.~\ref{fig:Sstar} we show values of $S^*$ used for each wavelength. 
We compared the frequency scaling of $S^{*}$ with that of a typical Spectral Energy Distribution
(SED) of submm galaxies SED$(\lambda) \propto \epsilon(\lambda) B(T,\lambda)$, redshifted to 
$z_s=3$ with emissivity 
$\epsilon(\lambda)=1-\exp[(-\lambda_0/\lambda)^{\beta}]$ and blackbody spectrum $B(T,\lambda)$ 
at temperature $T$. 
The values of $S^*$ are consistent with $\beta=1-2$ and $T=30-40$~$K$,
as expected  for high redshift submm galaxies.

The fit to the BLAST counts at $\lambda=500$, $350$ and $250$~$\mu$m falls further below the high flux measurement at shorter wavelengths, suggesting the need for a lower redshift 
population. Indeed, \cite{Ealetal09} have identified the radio and 24 $\mu$m 
counterparts of the bright BLAST sources.  
At 250 $\mu$m almost all of the bright sources are identified as being at 
redshifts lower than 1.  One third to one half of the sources in the highest flux 
bin at 500 $\mu$m come from sources with $z<1$. Removing these would lower 
the corresponding point in Fig.~\ref{fig:dndS_all}, in agreement with the model.
The remaining sources at 500 $\mu$m (and less than a tenth of the sources at $250$ $\mu$m)  
are likely the result of lensing.
Similar results should be expected with the upcoming release of the large-area 
Herschel surveys.

Since the lensed distributions at $z_s=3.0$ are consistent with SPT data points at 
both wavelengths, we study the range of magnifications and 
halo masses that would be contributing most in this case.  
As we consider larger values of $\Sobs$, in particular $\Sobs/S^*\simgt 10$, the 
observed sources come from intrinsically low flux sources which have been magnified 
significantly.   
Fig.~\ref{fig:PmuminSobs.vs.S} shows $P(\mumin|\Sobs)$ and $\langle \mu \rangle$ 
as a function of $\Sobs$  for SPT ($\lambda=1.4$ mm) and indicates magnifications 
that contribute most at each $\Sobs$.
For instance, for $\Sobs=20-40$~mJy, the right panel shows that $\langle \mu \rangle \sim 20-30$.
 Another way to see this is via $P(\mumin|\Sobs)$ (left panel), which integrates 
out the effects above a certain $\mumin$, and shows that for $\Sobs=20-40$~mJy, 
we have $P(\mumin|\Sobs)>0.5$ for $\mumin \sim 10-20$.

These results imply that magnifications of $10-30$ are necessary to explain the 
boost in $dn/dS$ at $\Sobs \sim 10~-~40$~mJy, if it is due to 
 lensing of an intrinsic Schechter distribution. 
Note that due to the finite size of submm galaxies, their
magnifications must have a cut-off, which has been estimated to be in the 
range $\mu \sim 10-40$ \citep{Peretal02} for galaxy lenses, and is
probably a factor of two or so larger for more massive lenses.
Indeed galaxies have  been measured with estimated magnifications of at
least $\sim 45$ \citep{PacScoCha08, Kneetal04}.

\begin{figure}
 \resizebox{88mm}{!}{\includegraphics[angle=0]{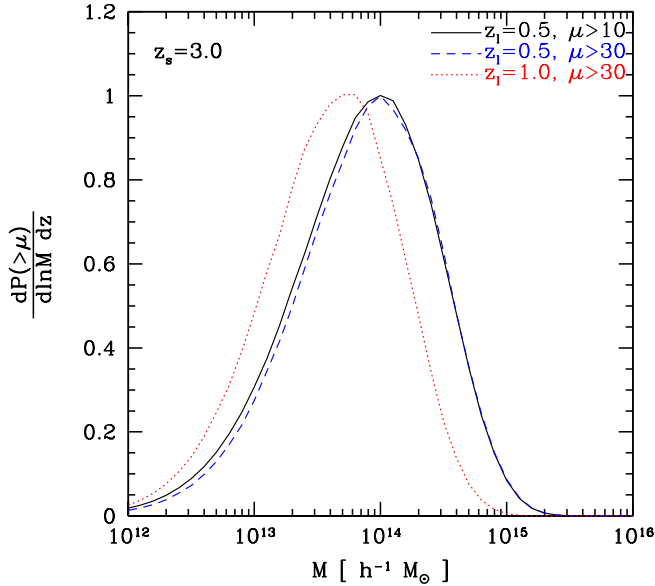}}
 \caption{ The integrand $d^2P(>\mu)/d\ln M dz_l$ as a function of halo mass $M$ for 
different values of $\mu$ and $z_l$. 
We have arbitrarily normalized these curves for better visualization. Note that the 
peak lensing contribution for source galaxies at $z_s=3$ comes from lens halos 
at $z_l\sim 0.5$. 
This does not include baryonic effects, which boost the contribution from 
$M\simlt 10^{13}h^{-1}\Msun$ halos.
}
 \label{fig:Integrand}
\end{figure}

In Fig.~\ref{fig:Integrand} we show $d^2P/d \ln M dz_l$ as a function
of halo mass for different values of $\mumin$ and $z_l$. This
indicates that halo masses above $10^{13}h^{-1}\Msun$ contribute
significantly to magnifications of $10-30$, with most of the
contribution coming from $\sim10^{14}h^{-1}\Msun$. Note that this 
does not include baryonic effects, which boost the contribution from lower mass 
($\simlt 10^{13}h^{-1}\Msun$) halos as discussed in Section 2.

%

We have assumed that all source galaxies are at a fixed redshift $z_s$, 
whereas in reality they have a redshift distribution which needs to 
be incorporated in the computation. 
The curves at different source redshifts shown in Fig.~\ref{fig:dndS_SPT_lens} 
provide approximate limits for what we can expect from a redshift distribution
of source galaxies. 
Our results imply that the bulk of the population of 
submillimeter galaxies are at redshifts $z\simgt 2$. 
The alternative explanation for the high flux measurements 
is to have a significant fraction of galaxies at  
very low $z$, which would easily have been observed in surveys such as
IRAS (indeed the SPT data shown remove the small fraction of such sources). 
And as shown in Fig.~\ref{fig:dndS_SPT_lens}, galaxies at higher
redshift but still at $z\simlt 1$ would not match the measurements as the 
magnification boost is insufficient (again within the context of an 
underlying Schechter distribution).

\section{Discussion}
\label{sec:discussion}

We have considered the possibility that bright millimeter and
submillimeter galaxy counts arise largely from a galaxy population at
high redshifts that is lensed by intervening galaxy groups and
clusters.  Our model predictions match the
counts from $500~\mu$m to $2~$mm from the BLAST, SCUBA,
AzTEC and SPT surveys (see Fig.~\ref{fig:dndS_all}).  
We find that the high flux 
SPT number counts can be explained by highly magnified galaxies from 
this high-$z$ population (once 
the known, low-$z$ counterparts detected in IRAS are removed). 
This high-$z$ galaxy population is described by a
Schechter luminosity function with $L^* \sim 2.5 \times
10^{12} L_\odot$ and a source redshift $z_s=3.0$.  Our model predictions fit 
the data for  $500~\mu$m~$< \lambda < 2~$mm by varying $S^*$ with 
wavelength within the range of typical submillimeter galaxy SEDs (see 
Fig.~\ref{fig:Sstar}). 

Our model has some simplifying assumptions, such as the fixed source redshift 
$z_s=3$, and there are significant measurement and theoretical uncertainties. 
A complete analysis would require a more detailed treatment of the lensing
and inclusion of the measured galaxy clustering. In addition, it has
been established that at the shorter wavelengths probed by BLAST, an
increasing fraction of sources lie at low-$z$ -- hence we can 
expect some smooth variation in the fraction of low-$z$ galaxies with observed
wavelength. 

Nevertheless our results in Fig.~\ref{fig:dndS_all} imply that 
current number counts do not require 
an additional galaxy population to explain the high flux measurements. 
Such a population has been invoked in theoretical models \citep[e.g.][]{Lagetal03} 
and suggested as a possible explanation (as well as lensing of the high-$z$ 
population) for the recent SPT measurements \citep{Vieetal09}. 
Indeed, adding a significant fraction of a second population could cause the 
predictions to exceed the measured counts once magnification effects for the 
high-$z$ population are included. 
It would also be difficult to explain how the scaling with wavelength 
of the high flux number counts is the same as the lower flux counts if they 
came from different galaxy populations.  Since lensing does not depend
on frequency, our model naturally follows this common scaling. 
  
Current SZ surveys with sensitivities of $3-7 \times 10^{14} h^{-1}\Msun$ 
cannot detect most halos that produce this lensing contribution. 
Conversely, looking for extremely bright objects in millimeter and
submillimeter wavelengths provides a way to find high-$z$
lensing halos associated with galaxy groups and clusters. Note that 
the number of halos that can be found in this way is only a small
fraction of all  halos within a given mass range. As discussed above, 
our model most likely underestimates the contribution from halos 
with $M\simlt 10^{13}\Msun$. It is therefore of great interest to 
investigate the lenses corresponding to the bright sources in current data. 

Follow up observations with optical telescopes should be able to
identify lensing group/cluster candidates up to $z\simeq 1$. These 
clusters and groups host Brightest Cluster Galaxies with luminosities 
$L\sim 10^{11}-10^{12} L_\odot$,  based on the low-$z$ results of
\cite{Johetal07} from the  Sloan Digital Sky Survey. Multi-band
optical imaging with limiting  
magnitude of $23-24$ (for the $r$ band) would enable identification 
of these groups and clusters. Conversely, targeted observations of
the high magnification regions of known strong lensing clusters could
provide detections of faint submm galaxies (which would
lie below the detection threshold without the magnification
boost). Similar to the use of clusters as gravitational telescopes in
optical imaging, this may also help resolve submm galaxies. Planned 
observations with AzTEC and the Large Millimeter Telescope have considered 
such an approach (David Hughes, private communication). 

\acknowledgements

We thank David Hughes, Roxana Lupu, Joaquin Vieira, Kim Scott, Ian Smail, Eric
Switzer and  
various members of the ACT collaboration for useful discussions. We benefited 
from discussions on strong lensing effects with Matthias Bartelmann, Gary Bernstein, 
Neal Dalal and Ravi Sheth. We are very grateful to Stefan Hilbert for sharing his simulation 
results and Eric Switzer for providing the SPT data. This work was
supported in part by an NSF-PIRE grant and AST-0607667.

\bibliographystyle{apj}
\bibliography{SubmmCounts}

\begin{thebibliography}{50}
\expandafter\ifx\csname natexlab\endcsname\relax\def\natexlab#1{#1}\fi

\bibitem[{{Aretxaga} {et~al.}(2007){Aretxaga}, {Hughes}, {Coppin}, {Mortier},
  {Wagg}, {Dunlop}, {Chapin}, {Eales}, {Gazta{\~n}aga}, {Halpern}, {Ivison},
  {van Kampen}, {Scott}, {Serjeant}, {Smail}, {Babbedge}, {Benson}, {Chapman},
  {Clements}, {Dunne}, {Dye}, {Farrah}, {Jarvis}, {Mann}, {Pope}, {Priddey},
  {Rawlings}, {Seigar}, {Silva}, {Simpson}, \& {Vaccari}}]{Areetal07}
{Aretxaga}, I., {et~al.} 2007, \mnras, 379, 1571

\bibitem[{{Austermann} {et~al.}(2010){Austermann}, {Dunlop}, {Perera}, {Scott},
  {Wilson}, {Aretxaga}, {Hughes}, {Almaini}, {Chapin}, {Chapman}, {Cirasuolo},
  {Clements}, {Coppin}, {Dunne}, {Dye}, {Eales}, {Egami}, {Farrah}, {Ferrusca},
  {Flynn}, {Haig}, {Halpern}, {Ibar}, {Ivison}, {van Kampen}, {Kang}, {Kim},
  {Lacey}, {Lowenthal}, {Mauskopf}, {McLure}, {Mortier}, {Negrello}, {Oliver},
  {Peacock}, {Pope}, {Rawlings}, {Rieke}, {Roseboom}, {Rowan-Robinson},
  {Scott}, {Serjeant}, {Smail}, {Swinbank}, {Stevens}, {Velazquez}, {Wagg}, \&
  {Yun}}]{Ausetal10}
{Austermann}, J.~E., {et~al.} 2010, \mnras, 401, 160

\bibitem[{{Barger} {et~al.}(1998){Barger}, {Cowie}, {Sanders}, {Fulton},
  {Taniguchi}, {Sato}, {Kawara}, \& {Okuda}}]{Baretal98}
{Barger}, A.~J., {Cowie}, L.~L., {Sanders}, D.~B., {Fulton}, E., {Taniguchi},
  Y., {Sato}, Y., {Kawara}, K., \& {Okuda}, H. 1998, \nat, 394, 248

\bibitem[{{Blain}(1996)}]{Bla96}
{Blain}, A.~W. 1996, \mnras, 283, 1340

\bibitem[{{Carlstrom} {et~al.}(2009){Carlstrom}, {Ade}, {Aird}, {Benson},
  {Bleem}, {Busetti}, {Chang}, {Chauvin}, {Cho}, {Crawford}, {Crites}, {Dobbs},
  {Halverson}, {Heimsath}, {Holzapfel}, {Hrubes}, {Joy}, {Keisler}, {Lanting},
  {Lee}, {Leitch}, {Leong}, {Lu}, {Lueker}, {McMahon}, {Mehl}, {Meyer}, {Mohr},
  {Montroy}, {Padin}, {Plagge}, {Pryke}, {Ruhl}, {Schaffer}, {Schwan},
  {Shirokoff}, {Spieler}, {Staniszewski}, {Stark}, \& {Vieira}}]{Caretal09}
{Carlstrom}, J.~E., {et~al.} 2009, ArXiv e-prints

\bibitem[{{Chapman} {et~al.}(2005){Chapman}, {Blain}, {Smail}, \&
  {Ivison}}]{Chaetal05}
{Chapman}, S.~C., {Blain}, A.~W., {Smail}, I., \& {Ivison}, R.~J. 2005, \apj,
  622, 772

\bibitem[{{Coppin} {et~al.}(2006)}]{Copetal06}
{Coppin}, K., {et~al.} 2006, \mnras, 372, 1621

\bibitem[{{Devlin} {et~al.}(2009)}]{Devetal09}
{Devlin}, M.~J., {et~al.} 2009, \nat, 458, 737

\bibitem[{{Dwek} {et~al.}(1998){Dwek}, {Arendt}, {Hauser}, {Fixsen}, {Kelsall},
  {Leisawitz}, {Pei}, {Wright}, {Mather}, {Moseley}, {Odegard}, {Shafer},
  {Silverberg}, \& {Weiland}}]{Dweetal98}
{Dwek}, E., {et~al.} 1998, \apj, 508, 106

\bibitem[{{Eales} {et~al.}(2009){Eales}, {Chapin}, {Devlin}, {Dye}, {Halpern},
  {Hughes}, {Marsden}, {Mauskopf}, {Moncelsi}, {Netterfield}, {Pascale},
  {Patanchon}, {Raymond}, {Rex}, {Scott}, {Semisch}, {Siana}, {Truch}, \&
  {Viero}}]{Ealetal09}
{Eales}, S., {et~al.} 2009, \apj, 707, 1779

\bibitem[{{Fisher} {et~al.}(1995){Fisher}, {Huchra}, {Strauss}, {Davis},
  {Yahil}, \& {Schlegel}}]{Fisetal95}
{Fisher}, K.~B., {Huchra}, J.~P., {Strauss}, M.~A., {Davis}, M., {Yahil}, A.,
  \& {Schlegel}, D. 1995, \apjs, 100, 69

\bibitem[{{Fixsen} {et~al.}(1996){Fixsen}, {Cheng}, {Gales}, {Mather},
  {Shafer}, \& {Wright}}]{Fixetal96}
{Fixsen}, D.~J., {Cheng}, E.~S., {Gales}, J.~M., {Mather}, J.~C., {Shafer},
  R.~A., \& {Wright}, E.~L. 1996, \apj, 473, 576

\bibitem[{{Fixsen} {et~al.}(1998){Fixsen}, {Dwek}, {Mather}, {Bennett}, \&
  {Shafer}}]{Fixetal98}
{Fixsen}, D.~J., {Dwek}, E., {Mather}, J.~C., {Bennett}, C.~L., \& {Shafer},
  R.~A. 1998, \apj, 508, 123

\bibitem[{{Fowler} {et~al.}(2010){Fowler}, {Acquaviva}, {Ade}, {Aguirre},
  {Amiri}, {Appel}, {Barrientos}, {Battistelli}, {Bond}, {Brown}, {Burger},
  {Chervenak}, {Das}, {Devlin}, {Dicker}, {Doriese}, {Dunkley}, {D{\"u}nner},
  {Essinger-Hileman}, {Fisher}, {Hajian}, {Halpern}, {Hasselfield},
  {Hern{\'a}ndez-Monteagudo}, {Hilton}, {Hilton}, {Hincks}, {Hlozek},
  {Huffenberger}, {Hughes}, {Hughes}, {Infante}, {Irwin}, {Jimenez}, {Juin},
  {Kaul}, {Klein}, {Kosowsky}, {Lau}, {Limon}, {Lin}, {Lupton}, {Marriage},
  {Marsden}, {Martocci}, {Mauskopf}, {Menanteau}, {Moodley}, {Moseley},
  {Netterfield}, {Niemack}, {Nolta}, {Page}, {Parker}, {Partridge}, {Quintana},
  {Reid}, {Sehgal}, {Sievers}, {Spergel}, {Staggs}, {Swetz}, {Switzer},
  {Thornton}, {Trac}, {Tucker}, {Verde}, {Warne}, {Wilson}, {Wollack}, \&
  {Zhao}}]{Fowetal10}
{Fowler}, J.~W., {et~al.} 2010, ArXiv e-prints

\bibitem[{{Gonzalez} {et~al.}(2009){Gonzalez}, {Clowe}, {Brada{\v c}},
  {Zaritsky}, {Jones}, \& {Markevitch}}]{Gonetal09}
{Gonzalez}, A.~H., {Clowe}, D., {Brada{\v c}}, M., {Zaritsky}, D., {Jones}, C.,
  \& {Markevitch}, M. 2009, \apj, 691, 525

\bibitem[{{Greve} {et~al.}(2004){Greve}, {Ivison}, {Bertoldi}, {Stevens},
  {Dunlop}, {Lutz}, \& {Carilli}}]{Greetal04}
{Greve}, T.~R., {Ivison}, R.~J., {Bertoldi}, F., {Stevens}, J.~A., {Dunlop},
  J.~S., {Lutz}, D., \& {Carilli}, C.~L. 2004, \mnras, 354, 779

\bibitem[{{Hilbert} {et~al.}(2007){Hilbert}, {White}, {Hartlap}, \&
  {Schneider}}]{Hiletal08}
{Hilbert}, S., {White}, S.~D.~M., {Hartlap}, J., \& {Schneider}, P. 2007,
  \mnras, 382, 121

\bibitem[{{Hilbert} {et~al.}(2008){Hilbert}, {White}, {Hartlap}, \&
  {Schneider}}]{Hiletal08b}
---. 2008, \mnras, 386, 1845

\bibitem[{{Hincks} {et~al.}(2008){Hincks}, {Ade}, {Allen}, {Amiri}, {Appel},
  {Battistelli}, {Burger}, {Chervenak}, {Dahlen}, {Denny}, {Devlin}, {Dicker},
  {Doriese}, {D{\"u}nner}, {Essinger-Hileman}, {Fisher}, {Fowler}, {Halpern},
  {Hargrave}, {Hasselfield}, {Hilton}, {Irwin}, {Jarosik}, {Kaul}, {Klein},
  {Lau}, {Limon}, {Lupton}, {Marriage}, {Martocci}, {Mauskopf}, {Moseley},
  {Netterfield}, {Niemack}, {Nolta}, {Page}, {Parker}, {Sederberg}, {Staggs},
  {Stryzak}, {Swetz}, {Switzer}, {Thornton}, {Tucker}, {Wollack}, \&
  {Zhao}}]{Hinetal08}
{Hincks}, A.~D., {et~al.} 2008, in Society of Photo-Optical Instrumentation
  Engineers (SPIE) Conference Series, Vol. 7020, Society of Photo-Optical
  Instrumentation Engineers (SPIE) Conference Series

\bibitem[{{Hughes} {et~al.}(1998){Hughes}, {Serjeant}, {Dunlop},
  {Rowan-Robinson}, {Blain}, {Mann}, {Ivison}, {Peacock}, {Efstathiou}, {Gear},
  {Oliver}, {Lawrence}, {Longair}, {Goldschmidt}, \& {Jenness}}]{Hugetal98}
{Hughes}, D.~H., {et~al.} 1998, \nat, 394, 241

\bibitem[{{Jain} \& {Lima}(2010)}]{JaiLim10}
{Jain}, B., \& {Lima}, M. 2010, ArXiv e-prints

\bibitem[{{Johnston} {et~al.}(2007){Johnston}, {Sheldon}, {Wechsler}, {Rozo},
  {Koester}, {Frieman}, {McKay}, {Evrard}, {Becker}, \& {Annis}}]{Johetal07}
{Johnston}, D.~E., {et~al.} 2007, ArXiv e-prints

\bibitem[{{Kneib} {et~al.}(2004){Kneib}, {van der Werf}, {Kraiberg Knudsen},
  {Smail}, {Blain}, {Frayer}, {Barnard}, \& {Ivison}}]{Kneetal04}
{Kneib}, J., {van der Werf}, P.~P., {Kraiberg Knudsen}, K., {Smail}, I.,
  {Blain}, A., {Frayer}, D., {Barnard}, V., \& {Ivison}, R. 2004, \mnras, 349,
  1211

\bibitem[{{Komatsu} {et~al.}(2009){Komatsu}, {Dunkley}, {Nolta}, {Bennett},
  {Gold}, {Hinshaw}, {Jarosik}, {Larson}, {Limon}, {Page}, {Spergel},
  {Halpern}, {Hill}, {Kogut}, {Meyer}, {Tucker}, {Weiland}, {Wollack}, \&
  {Wright}}]{Kometal09}
{Komatsu}, E., {et~al.} 2009, \apjs, 180, 330

\bibitem[{{Komatsu} {et~al.}(2010){Komatsu}, {Smith}, {Dunkley}, {Bennett},
  {Gold}, {Hinshaw}, {Jarosik}, {Larson}, {Nolta}, {Page}, {Spergel},
  {Halpern}, {Hill}, {Kogut}, {Limon}, {Meyer}, {Odegard}, {Tucker}, {Weiland},
  {Wollack}, \& {Wright}}]{Kometal10}
---. 2010, ArXiv e-prints

\bibitem[{{Lagache} {et~al.}(2003){Lagache}, {Dole}, \& {Puget}}]{Lagetal03}
{Lagache}, G., {Dole}, H., \& {Puget}, J. 2003, \mnras, 338, 555

\bibitem[{{Lagache} {et~al.}(2004){Lagache}, {Dole}, {Puget},
  {P{\'e}rez-Gonz{\'a}lez}, {Le Floc'h}, {Rieke}, {Papovich}, {Egami},
  {Alonso-Herrero}, {Engelbracht}, {Gordon}, {Misselt}, \&
  {Morrison}}]{Lagetal04}
{Lagache}, G., {et~al.} 2004, \apjs, 154, 112

\bibitem[{{Le Borgne} {et~al.}(2009){Le Borgne}, {Elbaz}, {Ocvirk}, \&
  {Pichon}}]{LeBetal09}
{Le Borgne}, D., {Elbaz}, D., {Ocvirk}, P., \& {Pichon}, C. 2009, \aap, 504,
  727

\bibitem[{{Li} {et~al.}(2005){Li}, {Mao}, {Jing}, {Bartelmann}, {Kang}, \&
  {Meneghetti}}]{Lietal05}
{Li}, G., {Mao}, S., {Jing}, Y.~P., {Bartelmann}, M., {Kang}, X., \&
  {Meneghetti}, M. 2005, \apj, 635, 795

\bibitem[{{Lima} {et~al.}(2009){Lima}, {Jain}, \& {Devlin}}]{LimJaiDev09}
{Lima}, M., {Jain}, B., \& {Devlin}, M. 2009, ArXiv e-prints

\bibitem[{{Lima} {et~al.}(2010)}]{Limetalinprep}
{Lima}, M., {et~al.} 2010, in prep.

\bibitem[{{Mason} {et~al.}(2006){Mason}, {Dicker}, {Korngut}, {Benford},
  {Devlin}, {Irwin}, {Moseley}, \& {MUSTANG collaboration}}]{Masetal06}
{Mason}, B.~S., {Dicker}, S., {Korngut}, P., {Benford}, D., {Devlin}, M.,
  {Irwin}, K., {Moseley}, H., \& {MUSTANG collaboration}. 2006, in Bulletin of
  the American Astronomical Society, Vol.~38, 1015--+

\bibitem[{{Micha{\l}owski} {et~al.}(2009){Micha{\l}owski}, {Hjorth}, \&
  {Watson}}]{Micetal09}
{Micha{\l}owski}, M.~J., {Hjorth}, J., \& {Watson}, D. 2009, ArXiv e-prints

\bibitem[{{Moshir} {et~al.}(1992){Moshir}, {Kopman}, \& {Conrow}}]{Mosetal92}
{Moshir}, M., {Kopman}, G., \& {Conrow}, T.~A.~O. 1992, {IRAS Faint Source
  Survey, Explanatory supplement version 2}, ed. {Moshir, M., Kopman, G., \&
  Conrow, T.~A.~O.}

\bibitem[{{Negrello} {et~al.}(2007){Negrello}, {Perrotta},
  {Gonz{\'a}lez-Nuevo}, {Silva}, {de Zotti}, {Granato}, {Baccigalupi}, \&
  {Danese}}]{Negetal07}
{Negrello}, M., {Perrotta}, F., {Gonz{\'a}lez-Nuevo}, J., {Silva}, L., {de
  Zotti}, G., {Granato}, G.~L., {Baccigalupi}, C., \& {Danese}, L. 2007,
  \mnras, 377, 1557

\bibitem[{{Oliver} {et~al.}(1996){Oliver}, {Rowan-Robinson}, {Broadhurst},
  {McMahon}, {Saunders}, {Taylor}, {Lawrence}, {Lonsdale}, {Hacking}, \&
  {Conrow}}]{Olietal96}
{Oliver}, S.~J., {et~al.} 1996, \mnras, 280, 673

\bibitem[{{Paciga} {et~al.}(2008){Paciga}, {Scott}, \& {Chapin}}]{PacScoCha08}
{Paciga}, G., {Scott}, D., \& {Chapin}, E.~L. 2008, ArXiv e-prints

\bibitem[{{Pearson} \& {Khan}(2009)}]{Peaetal09}
{Pearson}, C., \& {Khan}, S.~A. 2009, \mnras, 399, L11

\bibitem[{{P{\'e}rez-Gonz{\'a}lez} {et~al.}(2005){P{\'e}rez-Gonz{\'a}lez},
  {Rieke}, {Egami}, {Alonso-Herrero}, {Dole}, {Papovich}, {Blaylock}, {Jones},
  {Rieke}, {Rigby}, {Barmby}, {Fazio}, {Huang}, \& {Martin}}]{Peretal05}
{P{\'e}rez-Gonz{\'a}lez}, P.~G., {et~al.} 2005, \apj, 630, 82

\bibitem[{{Perrotta} {et~al.}(2002){Perrotta}, {Baccigalupi}, {Bartelmann}, {De
  Zotti}, \& {Granato}}]{Peretal02}
{Perrotta}, F., {Baccigalupi}, C., {Bartelmann}, M., {De Zotti}, G., \&
  {Granato}, G.~L. 2002, \mnras, 329, 445

\bibitem[{{Pope} {et~al.}(2006){Pope}, {Scott}, {Dickinson}, {Chary},
  {Morrison}, {Borys}, {Sajina}, {Alexander}, {Daddi}, {Frayer}, {MacDonald},
  \& {Stern}}]{Popetal06}
{Pope}, A., {et~al.} 2006, \mnras, 370, 1185

\bibitem[{{Rex} {et~al.}(2009){Rex}, {Ade}, {Aretxaga}, {Bock}, {Chapin},
  {Devlin}, {Dicker}, {Griffin}, {Gundersen}, {Halpern}, {Hargrave}, {Hughes},
  {Klein}, {Marsden}, {Martin}, {Mauskopf}, {Monta{\~n}a}, {Netterfield},
  {Olmi}, {Pascale}, {Patanchon}, {Scott}, {Semisch}, {Thomas}, {Truch},
  {Tucker}, {Tucker}, {Viero}, \& {Wiebe}}]{Rexetal09}
{Rex}, M., {et~al.} 2009, \apj, 703, 348

\bibitem[{{Schechter}(1976)}]{Sch76}
{Schechter}, P. 1976, \apj, 203, 297

\bibitem[{{Smail} {et~al.}(1997){Smail}, {Ivison}, \& {Blain}}]{SmaIviBla97}
{Smail}, I., {Ivison}, R.~J., \& {Blain}, A.~W. 1997, \apjl, 490, L5+

\bibitem[{{Smail} {et~al.}(2002){Smail}, {Ivison}, {Blain}, \&
  {Kneib}}]{Smaetal02}
{Smail}, I., {Ivison}, R.~J., {Blain}, A.~W., \& {Kneib}, J. 2002, \mnras, 331,
  495

\bibitem[{{Staniszewski} {et~al.}(2009){Staniszewski}, {Ade}, {Aird}, {Benson},
  {Bleem}, {Carlstrom}, {Chang}, {Cho}, {Crawford}, {Crites}, {de Haan},
  {Dobbs}, {Halverson}, {Holder}, {Holzapfel}, {Hrubes}, {Joy}, {Keisler},
  {Lanting}, {Lee}, {Leitch}, {Loehr}, {Lueker}, {McMahon}, {Mehl}, {Meyer},
  {Mohr}, {Montroy}, {Ngeow}, {Padin}, {Plagge}, {Pryke}, {Reichardt}, {Ruhl},
  {Schaffer}, {Shaw}, {Shirokoff}, {Spieler}, {Stalder}, {Stark},
  {Vanderlinde}, {Vieira}, {Zahn}, \& {Zenteno}}]{Staetal09}
{Staniszewski}, Z., {et~al.} 2009, \apj, 701, 32

\bibitem[{{Swinbank} {et~al.}(2010){Swinbank}, {Smail}, {Longmore}, {Harris},
  {Baker}, {De Breuck}, {Richard}, {Edge}, {Ivison}, {Blundell}, {Coppin},
  {Cox}, {Gurwell}, {Hainline}, {Krips}, {Lundgren}, {Neri}, {Siana}, {Stark},
  {Wilner}, \& {Younger}}]{Swietal10}
{Swinbank}, M., {et~al.} 2010, ArXiv e-prints

\bibitem[{{Vanderlinde} {et~al.}(2010){Vanderlinde}, {Crawford}, {de Haan},
  {Dudley}, {Shaw}, {Ade}, {Aird}, {Benson}, {Bleem}, {Brodwin}, {Carlstrom},
  {Chang}, {Crites}, {Desai}, {Dobbs}, {Foley}, {George}, {Gladders}, {Hall},
  {Halverson}, {High}, {Holder}, {Holzapfel}, {Hrubes}, {Joy}, {Keisler},
  {Knox}, {Lee}, {Leitch}, {Loehr}, {Lueker}, {Marrone}, {McMahon}, {Mehl},
  {Meyer}, {Mohr}, {Montroy}, {Ngeow}, {Padin}, {Plagge}, {Pryke}, {Reichardt},
  {Rest}, {Ruel}, {Ruhl}, {Schaffer}, {Shirokoff}, {Song}, {Spieler},
  {Stalder}, {Staniszewski}, {Stark}, {Stubbs}, {van Engelen}, {Vieira},
  {Williamson}, {Yang}, {Zahn}, \& {Zenteno}}]{Vanetal10}
{Vanderlinde}, K., {et~al.} 2010, ArXiv e-prints

\bibitem[{{Vieira} {et~al.}(2009){Vieira}, {Crawford}, {Switzer}, {Ade},
  {Aird}, {Ashby}, {Benson}, {Bleem}, {Brodwin}, {Carlstrom}, {Chang}, {Cho},
  {Crites}, {de Haan}, {Dobbs}, {Everett}, {George}, {Gladders}, {Hall},
  {Halverson}, {High}, {Holder}, {Holzapfel}, {Hrubes}, {Joy}, {Keisler},
  {Knox}, {Lee}, {Leitch}, {Lueker}, {Marrone}, {McIntyre}, {McMahon}, {Mehl},
  {Meyer}, {Mohr}, {Montroy}, {Padin}, {Plagge}, {Pryke}, {Reichardt}, {Ruhl},
  {Schaffer}, {Shaw}, {Shirokoff}, {Spieler}, {Stalder}, {Staniszewski},
  {Stark}, {Vanderlinde}, {Walsh}, {Williamson}, {Yang}, {Zahn}, \&
  {Zenteno}}]{Vieetal09}
{Vieira}, J.~D., {et~al.} 2009, ArXiv e-prints

\bibitem[{{Wilson} {et~al.}(2008){Wilson}, {Hughes}, {Aretxaga}, {Ezawa},
  {Austermann}, {Doyle}, {Ferrusca}, {Hern{\'a}ndez-Curiel}, {Kawabe},
  {Kitayama}, {Kohno}, {Kuboi}, {Matsuo}, {Mauskopf}, {Murakoshi},
  {Monta{\~n}a}, {Natarajan}, {Oshima}, {Ota}, {Perera}, {Rand}, {Scott},
  {Tanaka}, {Tsuboi}, {Williams}, {Yamaguchi}, \& {Yun}}]{Wiletal08}
{Wilson}, G.~W., {et~al.} 2008, \mnras, 390, 1061

\end{thebibliography}

\end{document}